\documentclass{pnastwo}

\url{-}

\usepackage{graphicx} 
\usepackage{subfigure}
\usepackage{bm} 
\graphicspath{}
\usepackage{subfloat}
\usepackage{longtable}
\usepackage{graphics}

\newcommand{\width}{0.47\textwidth}

\begin{document}

\title{New x-ray measurements in Helium-like Atoms increase discrepancy between experiment and theoretical QED}

\author{Christopher T Chantler\affil{1}{The University of Melbourne, Melbourne, Australia},
Andrew T Payne\affil{1}{}, John D Gillaspy\affil{2}{National Institute of Standards and Technology, 100
Bureau Drive, Gaithersburg, MD 20899, USA}, Lawrence T Hudson\affil{2}{}, Lucas F Smale\affil{1}{}, Albert Henins\affil{2}{}, Justin A Kimpton\affil{3}{Australian Synchrotron, Melbourne, Victoria 3000, Australia}, Endre Takacs\affil{4}{Department of Physics and Astronomy, Clemson University, Clemson, South Carolina 29634 USA}\affil{5}{Experimental Physics Department, University of Debrecen, Bem ter 18/a, Debrecen, Hungary H-4026}}

\contributor{}

\significancetext{Quantum Electro-Dynamics (QED) is the best tested theory of our physical world, yet significant discrepancies have emerged between attempts to apply QED to the simplest atomic systems. The pattern of discrepancy observed in Helium-like systems is determined to fit a functional form corresponding to one or more Feynman diagrams with high significance. We investigate experiments with absolute uncertainties. The consequences on our understanding are discussed.}

\maketitle

\begin{article}
\begin{abstract}
{A recent 15 parts-per-million (ppm) experiment on muonic hydrogen ($p^+ \mu^-$)
found a major discrepancy with 
QED and independent nuclear size determinations.
Here we find a significant discrepancy 
in a different type of exotic atom, a medium-Z nucleus with two electrons.
Investigation of the data collected is able to discriminate between available QED formulations and reveals a pattern of discrepancy of almost 6 standard errors of experimental results from the most recent theoretical predictions with a functional dependence proportional to $Z^n$ where $n\simeq 4$.
In both the muonic and highly charged systems, the sign of the discrepancy is the same,
with the measured transition energy higher than predicted.
Some consequences are possible or probable, and some are more speculative.
This may give insight into effective nuclear radii, the Rydberg, the fine-structure constant or unexpectedly large QED terms.
}
\end{abstract}


\abbreviations{QED, Quantum Electrodynamics; EBIT, Electron Beam Ion Trap; ppm, parts-per-million; s.e., standard error}

\dropcap{I}n this article we study the pattern of discrepancy between two-electron experiment and theory for X-ray transitions to core holes. Recent work has raised this area as one of the current anomalies in QED computation.
Quantum electrodynamics (QED) is one of the most important foundations of modern physics. 
The five standard deviation inconsistency between a 15 ppm (parts-per-million) measurement of a muonic hydrogen transition and theory~\cite{PohlNature} has led to four years of intensive research by many groups around the world.
Leading theorists consider the discrepancy of 0.42 meV to be well outside possible causes within the Standard Model, claimed to have an uncertainty of no more than $\pm$0.01 meV~\cite{Jentschura2011}

This puzzling situation has stimulated much theoretical activity and highlights the current difficulty in low-\emph{Z} atomic spectroscopy due to complexities of the nucleus. In precision measurements with atomic hydrogen~\cite{Udem}, progress is stalled by uncertainties in nuclear form factors and nuclear polarization which render the last few digits of available experimental accuracy underutilized.  At high-\emph{Z}, the strong enhancement of nuclear interactions also limits the degree to which available experimental measurements can be used to test QED~\cite{chantler_kimpton}.
For medium-\emph{Z}, nuclear uncertainties do not limit the interpretation of atomic spectroscopy. 
The overall magnitude of the contribution from nuclear size and shape is small, and the uncertainty on the magnitude is smaller still:  for the case of titanium, the uncertainty in the nuclear radius even in early atomic structure calculations is 0.012 fm~\cite{johnson_soff}, less than half that of the proton~\cite{johnson_soff} and much less than the 0.42 fm discrepancy from the muonic hydrogen experiment. 

Spectroscopy of highly charged ions and muonic atoms probe a relatively unexplored regime of physics in which the peak of the radial wavefunction of the lepton is reduced by more than an order of magnitude.  Effects associated with QED and the nucleus are greatly enhanced, due to the increased overlap of the nucleus with the wavefunction of the orbiting lepton.  In the case of muonic hydrogen, the lepton orbital radius is reduced by the mass of the lepton, while in the case of highly charged ions the lepton orbital radius is decreased by the increased nuclear charge. While hydrogenic (1-electron) atomic systems are exotic and are critical challenges for theory and experiment, helium-like (He-like) atomic systems lie at one of the forefronts of QED research~\cite{Artemyev}, because they display qualitatively new effects (including the `two-electron Lamb shift') which are not present at any level in one-electron ions. 

Crucial higher order 1e-QED terms scale as $\alpha^2(Z\alpha)^6$ and $\alpha (Z\alpha)^7$for hydrogen, while similar 2e-QED terms scale as $\alpha^2(\emph{Z}\alpha)^{6}$ and $\alpha^2(\emph{Z}\alpha)^{7}$. Hence an increase in $Z$ results in a dramatic increase in the magnitude of higher-order contributions \cite{Karshenboim_Ivanov}. 
If the $Z$-expansions used for hydrogen remained valid, then a 1 ppm measurement in the middle of the periodic table would be equivalent to testing some higher-order terms in hydrogen to a few parts in $10^{15}$~\cite{chantler_kimpton}.  
Moreover, in the mid-$Z$ range, some of the perturbative expansions fail so that non-perturbative methods are required.  The mid-$Z$ crossover range between neutral atoms and very high-$Z$ few-electron ions is the focus of this paper. Tests with atoms and muonic hydrogen have focussed upon the proton radius, while the real change may lie elsewhere.

Precision measurements of X-ray energies require high 
resolution and must be undertaken using wavelength dispersive
spectroscopy employing Bragg diffraction. Tests of QED in helium-like
systems often observe the diffraction profile of 
\emph{w}(1s$^2$($^1$S$_0$)$\rightarrow$1s2p($^1$P$_1$)), and less frequently the
\emph{x}(1s$^2$($^1$S$_0$)$\rightarrow$1s2p($^3$P$_2$)),
\emph{y}(1s$^2$($^1$S$_0$)$\rightarrow$1s2p($^3$P$_1$)) and
\emph{z}(1s$^2$($^1$S$_0$)$\rightarrow$1s2s($^3$S$_1$)) transitions. 
\section{Inconsistency with the null hypothesis}
Taken individually, recently measured helium-like transition energies of titanium demonstrate significant deviation from the most recent comprehensive ab initio theoretical QED formulation~\cite{Artemyev}, with significances of 2.9 s.e. for the $w$-line~\cite{Chantler2012PRL}, 1.4 s.e. for the $z$-line, and 1.3 s.e. and 0.3 s.e. for the $y$- and $x$-lines respectively~\cite{PayneJPB2014}.  All lie higher than the predictions of theory. The weighted mean of these four deviations from theory is non-zero by 3.2 s.e.

When all the assessed literature claiming absolute measurements for the four lines are taken together
and the weighted mean is calculated for each value of $Z>15$, then compared to predictions~\cite{Artemyev}, the $\chi^2_r$ is found to be 3.5 (for no fit parameters), corresponding to a probability of less than $P=1.2\times 10^{-5}$ that the data are adequately described by the predictions.  This presents a challenging puzzle as the uncertainty on the predictions are reported to be much smaller than even the most precise experiments and the results of four independent calculations~\cite{Artemyev,Cheng,Drake88,Plante} by leading groups agree with each other to within a fraction of the observed deviation from experiment.  
Some 18 measurements are added for $x$, $y$ and $z$ transitions, including our recent data.
The possibility that the experiments suffer from a common systematic error seems unlikely given that this set includes 63 results distributed over 28 different experiments from at least 12 different groups, over more than four decades of time. In addition, the independent measurements at each value of $Z$ are in good agreement with their weighted means.

There are other measurements in the literature at lower values of $Z$, not directly comparable to~\cite{Artemyev} because that work is only tabulated down to $Z=12$ and becomes increasingly uncertain as $Z$ is reduced.  In fact, the difference in calculated energy of the ground level with respect to a competing calculation crosses over and changes sign around $Z=15$.

\section{Pattern of discrepancy}
\begin{figure}\begin{center}\includegraphics[width=\width]{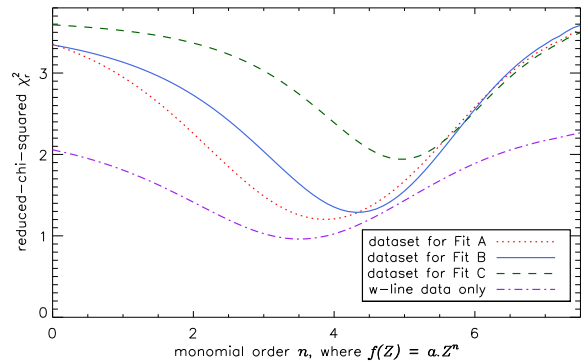} 
\caption{A persistent discrepancy. Plot of $\chi^2_r$ by monomial order for various data sets combining literature values with our experimental results.The dot-dash curve (magenta) includes only measurements of the $w$-line~\cite{ReplyPRL2013}; the red dotted line is Fit `A', including all absolute $w$, $x$, $y$ and $z$ transition measurements with claimed uncertainties above 10 ppm; the solid blue line represents the $\chi^2_r$ valley including~\cite{Amaro2012}; and the green dashed line includes both~\cite{Amaro2012,Kubicek2012}.
the dotted line includes $w$, $x$, $y$ and $z$ transitions.
\label{fig:pnas_figure2}}
\end{center}
\end{figure}

Our earlier work analyzed the strongest of the four lines and included data from the literature up through the end of 2011, and suggested a systematic discrepancy from theory with a functional dependence proportional to $Z^3$~\cite{Chantler2012PRL}. A later investigation demonstrated that the consideration of non-integer exponents led to a minimum at $Z^{3.5}$~\cite{ReplyPRL2013} (lower curve in Fig.\ref{fig:pnas_figure2}). 
Quadrupling the number of spectral peaks does much more than increase the statistical significance of the deviation, it rules out a range of systematics relating to satellites and line shifts from adjacent peak overlap, for example, which may be unrecognised. The consistent pattern of discrepancy that results when all four lines are included also suggests that the problem lies primarily in the calculation of the ground level (common to all four lines) or that the error in most of the upper levels is of the same magnitude for each value of $Z$.

\begin{figure}
\begin{center}
\includegraphics[width=0.5\textwidth]{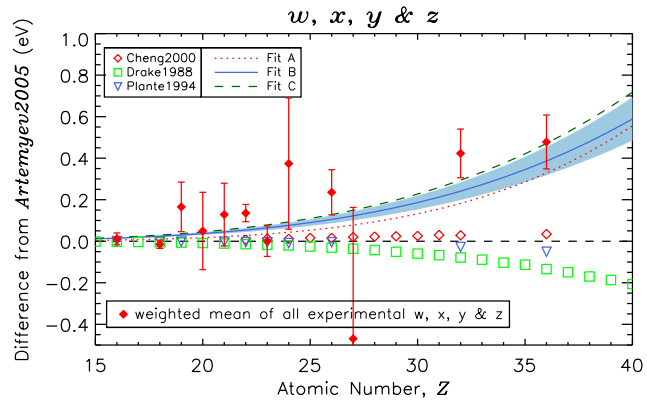} 
\caption{\label{subfigures} Discrepancy of experimental data from latest theory for two-electron systems. Difference of weighted mean results for \emph{w,x,y,z} transitions. 
The [red] dotted line is {Fit `A'} (see text and Table \ref{table_data};
 $Z^4$ dependence, $\chi^2_r=1.2$); the [blue] solid line is {Fit `B'} (including~\cite{Amaro2012} and assuming $Z^4$ dependence, with one s.e. shading - (68\%) confidence interval around the fit to demonstrate consistency with the other fits within uncertainty, $\chi^2_r=1.3$); the [green] dashed line is {Fit `C'} including~\cite{Amaro2012,Kubicek2012}, scaling as $Z^5$; all are fitted across the range $Z$ $\in [15,92]$. The error bar shown on each point is the s.e. 
Experimental results for the $w,x,y,z$ lines included in Fit `A' are plotted as weighted means for each $Z$ (Table \ref{table_dataWM}). 
Theoretical formulations presented, relative to Artemyev et al.~\cite{Artemyev} are Cheng~\cite{Cheng}, Drake~\cite{Drake88} and Plante~\cite{Plante}.}
\end{center}
\end{figure}

\begin{figure}
\begin{center}
\includegraphics[width=0.5\textwidth]{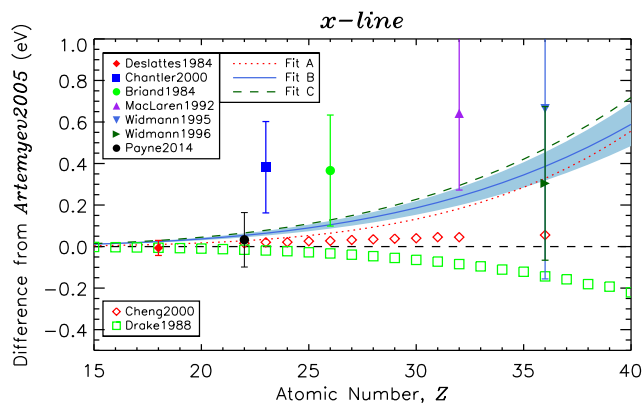} 
\caption{\label{subfiguresx} Discrepancy of experimental data from latest theory for two-electron systems for \emph{x} $(1s2p ^3P_2 \to  {1s^2} ^1S_0)$, plotted across mid-$Z$ ($Z$ $\in [15,40]$),  from Artemyev et al.~\cite{Artemyev}.  
Lines as per Fig~\ref{subfigures}.
Experimental results for the $x$ lines included in Fit `A' are plotted for each $Z$. 
The result of our recent work is plotted as a black circle.Theoretical formulations presented, relative to Artemyev et al.~\cite{Artemyev} are Cheng~\cite{Cheng}, Drake~\cite{Drake88} and Plante~\cite{Plante}.
None of these different advanced computations are consistent with the experimental data.
Although the result is dominated by the statistical uncertainty for this transition, there is very good agreement with the imputed discrepancy.
Experimental results plotted are~\cite{PayneJPB2014,DeslattesBeyer,CTC00,Briand1984,MacLaren1992,Widmann,Widmann1996}
}
\end{center}
\end{figure}

\begin{figure}
\begin{center}
\includegraphics[width=0.5\textwidth]{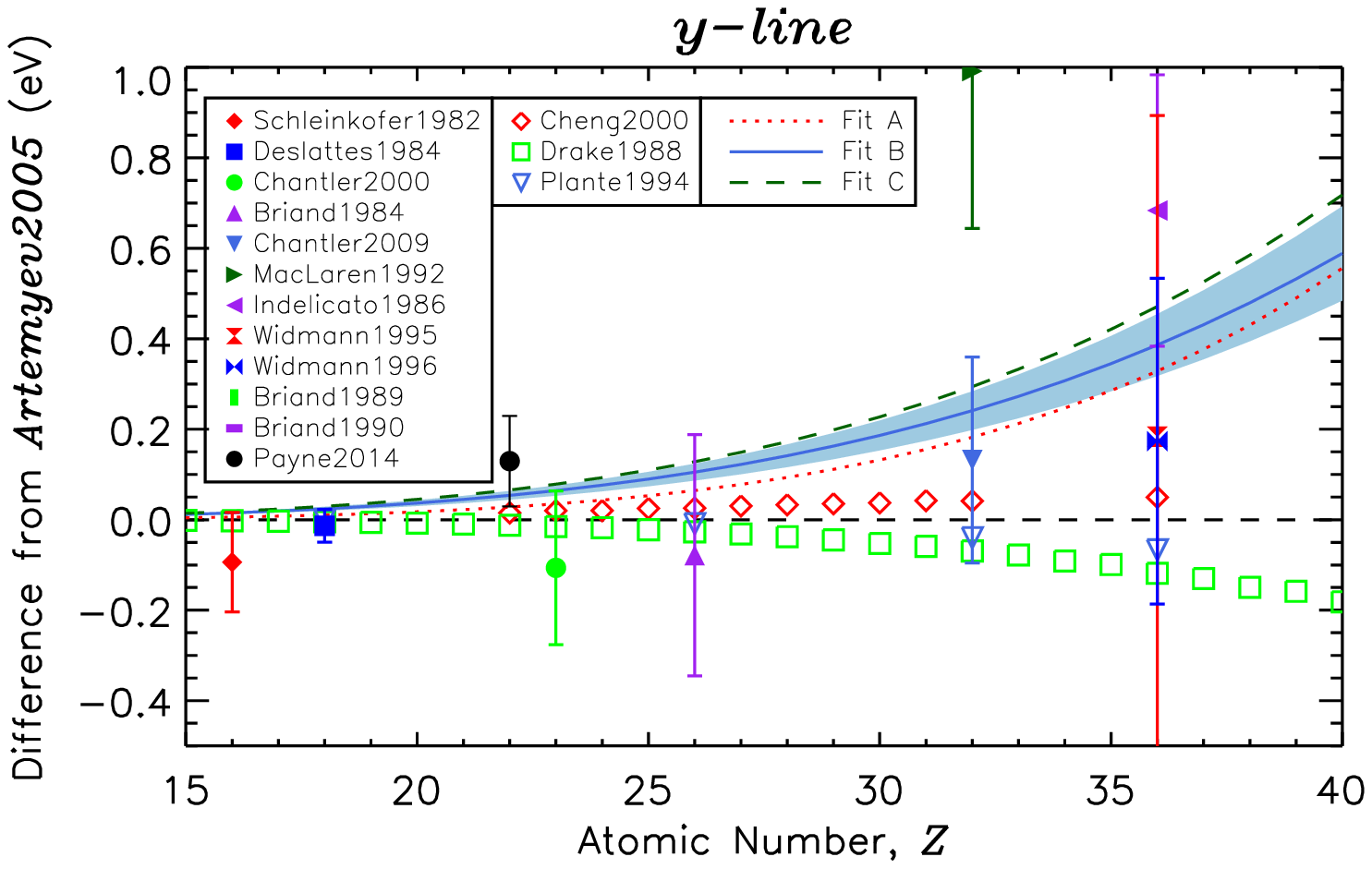} 
\caption{\label{subfiguresy} Discrepancy of experimental data from latest theory for two-electron systems for \emph{y} ($1s2p ^3P_1 \to  {1s^2} ^1S_0$) as per Fig.~\ref{subfiguresx}. 
Lines as per Fig~\ref{subfigures}.
Experimental results plotted are~\cite{PayneJPB2014,DeslattesBeyer,CTC00,Germanium,schleinkofer,Briand1984,Briand1989,Briand1990,MacLaren1992,Indelicato86,Widmann,Widmann1996}.
}
\end{center}
\end{figure}

\begin{figure}
\begin{center}
\includegraphics[width=0.5\textwidth]{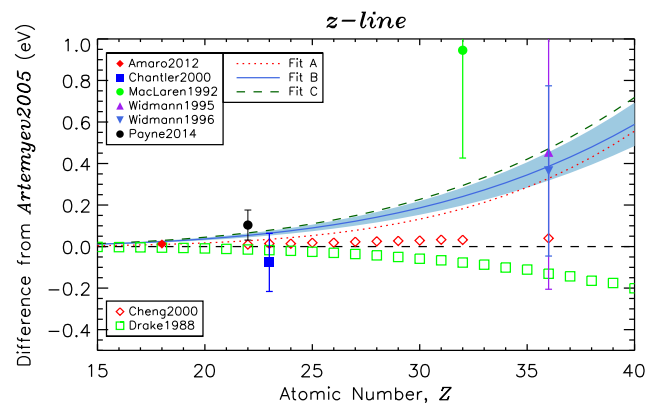} 
\caption{\label{subfiguresz} Discrepancy of experimental data from latest theory for two-electron systems for \emph{z} ($1s2s ^3S_1 \to  {1s^2} ^1S_0$) as per Fig.~\ref{subfiguresx}.
Lines as per Fig~\ref{subfigures}.
Experimental results plotted are~\cite{PayneJPB2014,CTC00,MacLaren1992,Widmann,Widmann1996}.
}
\end{center}
\end{figure}

Deviations of the experiments from theory for all four lines are plotted as a function of $Z$ (Figs~\ref{subfigures},\ref{subfiguresx},\ref{subfiguresy},\ref{subfiguresz}), and are statistically consistent with a single monomial, rising approximately as $Z^4$.  
Theory predicts that the first order QED corrections (vacuum polarization and self-energy) should scale as approximately $Z^4$ as well.  
Specific Feynman diagrams involving virtual photons and electron correlation follow expansions of order $Z^4$, $Z^6 (1/Z)$, $Z^6 (1/Z)^2$ and $Z^4(1/Z)$, for example.
The current level of accuracy of the experiments do not warrant fitting a polynomial of more than one degree of freedom, hence we consider only the coefficient of a monomial of order $n$, where $n$ is taken to be the constant value which minimizes the $\chi^2_r$ of the deviations of the weighted means at each value of $Z$ from the predictions of theory.
The consideration of a high-order polynomial is self-evidently consistent with Feynman diagram expectation values and correlation orders. The null hypothesis (i.e. a constant offset discrepancy from theory with a zero amplitude) has the same number of fitting coefficients and appears improbable, so we are comparing like with like in the statistical evaluation.

To check the possibility that one or more experimental approaches might have a recurring systematic error we have performed a variety of robustness tests by systematically deleting all of the data from one or more types of experimental arrangements (those with the most data; with the smallest error bars; etc.) and refit. The data set composed of our results and results prior to 2012 is robust against this test.  We also grouped these data by subset of the photon source and we find robust evidence for a strong $Z$-dependent deviation from theory. For example, looking at only EBIT data, since that subset is believed to suffer less from systematic corrections, we find a $Z^{3.2}$ dependence of discrepancy, based on the $w$-line only, with a $\chi^2_r = 1.06$. Inclusion of EBIT data from the literature on $x$, $y$ and $z$ yields the same optimised dependence of discrepancy ($Z^{3.2}$) and 
a significance of 4.6 (i.e. a 4.6 standard error discrepancy from theory).
Furthermore, the evidence suggests that the measurements in the literature up through 2011 have reasonably accurately estimated their errors, and can be treated as statistically independent.

We now consider the effect of two recent measurements in argon~\cite{Amaro2012,Kubicek2012} which appear to fall into a different category from earlier data. These two measurements claim an uncertainty at the level of nearly one part-per-million (ppm), about an order of magnitude better than the best measurements performed at any other value of Z, and hence bring with them the possibility of significantly skewing the results of a global fit.

The primary conclusion of a $Z$-dependent discrepancy remains unchanged whether or not these two recent data points are included in the analysis. The statistical consistency of the global deviation fit however can be significantly affected.  To investigate this, we present the results of three different fits to subsets of the
data:  Fit `A' including~\cite{Chantler2012PRL,PayneJPB2014,Aglitskii1974,Aglitsky,Cocke1974,Kubicek2009,schleinkofer,Briand1983,Bruhns2007,DeslattesBeyer,Dohmann1979,Neupert1971,Beiersdorfer89,CTC00,Briand1984,CTC07,Germanium,MacLaren1992,Indelicato86,Widmann,Widmann1996,Briand1989,Thorn2009,Briand1990}, Fit `B' including~\cite{Amaro2012}, and Fit `C' including~\cite{Amaro2012,Kubicek2012}.

We obtain a discrepancy given by Fit `A', 
$\Delta E  = (2.8\times 10^{-7}\pm 4.9\times 10^{-8})Z^4$ eV,  $\chi^2_r$=1.2. The uncertainty of the fit coefficient represents a 5.74 s.e. deviation from zero. 
Such a discrepancy would occur in a normal distribution with a probability of only $5\times 10^{-9}$ [5.74 s.e.]. 
This fit represents some 61 spectral lines, 
and lies within one s.e. of the result based on the $w$-line only.
This fit is presented in Figs~ \ref{subfigures},\ref{subfiguresx},\ref{subfiguresy},\ref{subfiguresz}, presenting individual data points for the $x$, $y$ and $z$ lines respectively, indicating the common pattern with increasing atomic number. 
All data are plotted against the theoretical predictions of Artemyev et al.~\cite{Artemyev} which has been the benchmark standard for helium-like QED theory for a number of years.
Three of the most highly referenced theoretical predictions from 1988~\cite{Drake88}, 1994~\cite{Plante}, 2000~\cite{Chen} and the current standard of 2005~\cite{Artemyev} are clustered together, indicating the advance and convergence of scholarship.

Amaro et al.~\cite{Amaro2012} do not report the $w$-transition because the state-population mechanisms of the ECRIS source rendered it relatively weak.  Their reported $z$-transition is discrepant from theory with the same sign as our results.
Including~\cite{Amaro2012} in the overall fit yields two results with the same goodness-of-fit: (1)
$\Delta E  = (4.2\times 10^{-8}\pm 7.4\times 10^{-9})Z^{4.5}$ eV,  $\chi^2_r$=1.3, a 5.7 s.e.~deviation of the coefficient from zero; and (2) Fit `B' (Figs~ \ref{subfigures},\ref{subfiguresx},\ref{subfiguresy},\ref{subfiguresz}), with
$\Delta E  = (2.3\times 10^{-7}\pm 4.0\times 10^{-8})Z^{4}$ eV,  $\chi^2_r$=1.3, also a 5.7 s.e.~deviation from zero. The deep valley of the $\chi^2_r$ surface (Fig.~\ref{fig:pnas_figure2}, `w-line' or Fits `A' or `B') argues strongly for a discrepancy and that a power law with $n\simeq 4$ (to within an uncertainty of about $\pm 0.5$) is the best fit. The $\chi^2_r$ remains consistent with unity, arguing for a self-consistent data set. The consistent patterns of discrepancy can be seen in the plots.
The power law dependence is robust and significant in all variations. Fits (1) and (2) (`B') lie within 1 s.e.~of Fit `A'.

There has been much confusion about the status of experiments at $Z=18$. Our current Fit `A' predicts at $Z=18$ a deviation from theory of $+0.029(5)$ eV, while~\cite{Amaro2012} found $+0.012(9)$ eV ($\pm 0.0077$ eV without inclusion in quadrature of claimed theoretical uncertainty). This~\cite{Amaro2012} is within two s.e. of both our predicted result and the previous theory, despite its low claimed uncertainty of 2.5 ppm (though of course in a different experimental regime). 
This measurement is a high-precision result using a powerful ion source (ECRIS), a new methodology, and discussion of a range of systematics.
The paper reports a $z$-transition so could not be discussed in terms of earlier analysis of the $w$ line consistency, but can now be discussed in relation to a common trend found in the four lines.
The inclusion or exclusion of \cite{Amaro2012} does not significantly impact upon the primary or numerical conclusions, nor the magnitude (number of standard errors) of the discrepancy. Specifically, with or without this data point, the discrepancy is 5.7 standard errors. The point also lies within the normal distribution of measured points relative to either Fit A or Fit B.


Including a second recent measurement~\cite{Kubicek2012} in the overall fit yields Fit `C', also plotted in the Figures, $\Delta E  =$ $\left(5.4\times 10^{-9}\pm 1.1\times 10^{-9}\right)Z^5$ eV,  $\chi^2_r=1.9$. This remains a 4.8 s.e.~deviation from current theory, within about one s.e.~of Fit `B'.
It~\cite{Kubicek2012} does skew the optimal order, the significance is only marginally weakened and the clear pattern of discrepancy from theory remains manifest under all permutations. This result is still inconsistent with theory with a probability of only $8\times 10^{-7}$ 
(4.8 s.e.~using normal distributions). 
According to~\cite{Amaro2012}, the authors of~\cite{Kubicek2012} neglected at least one potentially large 5 ppm shift especially compared to their claimed uncertainty of 1.5 ppm. 
Further, this accuracy implies that their published correction for over 60 ppm of line curvature has been corrected to an accuracy of approximately 1 ppm.


Table \ref{table_dataWM} lists all of the weighted means of the data included in our analysis.  
Fit `A' included all data cited except~\cite{Amaro2012,Kubicek2012}; Fit `B' included~\cite{Amaro2012}; and Fit `C' included~\cite{Amaro2012,Kubicek2012} (Table \ref{table_data}).
From ~\cite{Aglitsky}, we include only the subset of their data that they designate as arising from ``direct measurements" of the satellite correction normalization ratios ($Z<28$); if we include their higher-Z results into our analysis, it does not significantly change our conclusions, but it does entangle the measurement with complex theoretical modelling untested in this regime.  
For the measurement of Bruhns et al.~\cite{Bruhns2007}, we use their claimed absolute uncertainty (rather than their uncertainty relative to calculated values for calibration lines in other ions) for direct comparison with our present results and with other claimed absolute measurements. 
The consistency of the pattern of discrepancy is remarkable - the discrepancies lie on one side of theory -  and this consistency across multiple spectral lines leads to the apparent universality of the fit.

\begin{figure}
\begin{center}
\includegraphics[width=0.5\textwidth]{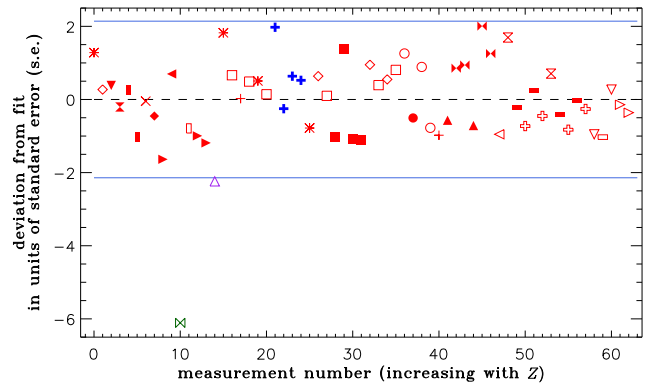} \caption{Deviation from `Fit A' 
of the data published up through~\cite{Kubicek2012}, in units of the published experimental error bars. \label{fig:science_outliera}}
\end{center}
\end{figure}

The very recent result~\cite{Kubicek2012} stands apart, in part by virtue of its extremely small claimed uncertainty (Fig.~\ref{fig:science_outliera}).
This datum would become consistent with the overall data set if the reported uncertainty were some 3 to 6 times the claimed value. 
Without such an expanded uncertainty, the probability that this point is consistent with the fit of the other points is $P < 10^{-9}$ based upon the 6.1 s.e.~discrepancy.

The body of experimental literature, arising from many different groups across a range of elements and using quite different experimental excitation and detection methods, is consistent with the approximately $Z^4$ deviation that we report here. Other authors have considered experimental measurements of QED to deviate from theory in past work
or have raised questions about specific values of $Z$~\cite{PohlNature,Chantler2004,Beiersdorfer89,chantlerthesis}, 
but these conjectures have sometimes been reversed in subsequent reports~\cite{Widmann1996,CTC00} reflecting the scarcity of all-Z data and large uncertainties at the time.
Here, both by increasing the accuracy of our own measurement and by performing a meta-analysis of the global data set, we postulate and present statistically strong and systematically consistent evidence for a functional dependence which may relate to particular correlation and Feynman diagrams.

The titanium measurements are the first set of He-like ions in which a highly significant discrepancy
has withstood the comparison across multiple spectral lines. 
Previously investigations typically either have had insufficient signal-to-noise,
too narrow a spectral bandpass, or inadequate calibration accuracy necessary to show such
a pattern of internal consistency. By including extensive data sets from the earlier literature, we minimise the possibility of a specific experimental systematic. By observing and including several lines we limit the possibility of any  asymmetry or line blends from impurities shifting one line in a particular direction.

\section{Harmonizing theory and experiment}

At present, QED treatment of low-$Z$ and high-$Z$ few-electron systems are undertaken with significantly different starting points and mathematical techniques. This work bridges the two regimes in order to stimulate the development of an improved universal computational methodology. Establishing this is desirable from a fundamental perspective, and crucial for the reliability of a wide range of practical applications. These include new classes of calibration standards based upon hydrogenic and helium-like energy levels~\cite{Amaro2012,anagnostopoulos2} and the multiplicity of novel laser techniques in high-field and high-energy-density applications~\cite{Wabnitz}. These experimental probes of the quantum vacuum provide the foundations for the determination of the current values of the constants of nature including the Rydberg constant~\cite{Weidemuller} and the fine structure constant~\cite{Flambaum1999}. Finally, this work calls into question the degree to which such atomic physics understanding has converged, and points, like the muonic hydrogen work, to further critical inquiry.

One interpretation of the fact that the same universal fit seems to be consistent for the patterns of discrepancy for $w,x,y$ and $z$ transitions is that the discrepancy could lie in the computation of the 1s$^2$ ground state energies and Lamb shift. The evidence is at least suggestive of a common or similar discrepancy for  $z$ (1s2s $^3S_1 \to $ 1s$^2$ $^1S_0$) and $w$ (1s2p $^1P_1 \to $ 1s$^2$ $^1S_0$), but there are few measurements of the $z$ transition to the accuracy required.

\begin{figure}
\begin{center}
\includegraphics[width=\width]{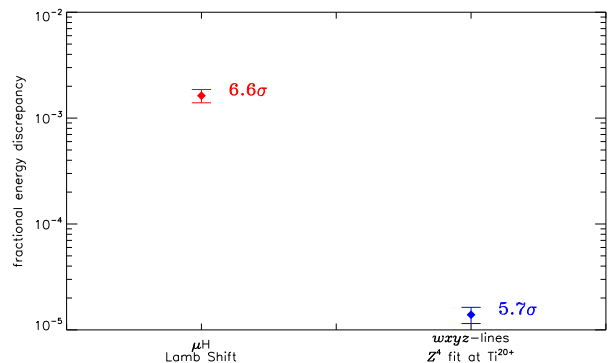}
\end{center}
\caption{Comparison of current status of two key discrepancies of QED-sensitive measurements from accepted energy determinations, divided by the magnitude of the energy measured. The scales in different types of experiments are quite different. 
The error bar shown on each point is the standard error of the measurement.  The label gives the ratio of the discrepancy to the uncertainty.
The muonic hydrogen Lamb shift energy~\cite{Science2013} is compared to the result using the CODATA rms proton charge radius~\cite{CODATA2006}.
The He-like \emph{wxyz}-transition discrepancy is determined as the transition energy difference of our results from the theory of Artemyev~\cite{Artemyev}, as a fraction of that theoretical energy at $Z=22$.\label{fig:nature_figure3Ea}
}
\end{figure}

In this case, we could have once again the possibility of a form factor, effective radius or some such similar interaction as has been proposed for the muonic hydrogen discrepancy. 
Indeed, a comparison of the magnitude of the muonic hydrogen discrepancy with the current status for helium-like medium-$Z$ systems is intriguing (Fig.~\ref{fig:nature_figure3Ea}). However, extensive theoretical investigation of that anomaly has thus far found no explanation based upon the effective radius, polarisation of the nucleus or shape.

If the 1s orbital is responsible, it does not prove an error of one-electron QED, which has been attested in experiments for the one-loop terms. It could relate to an interplay between the 1s electrons and their correlation energies. The former can be represented by $Z^4$ terms; the latter by $Z^3$ or $Z^5$.
The various theoretical predictions consider their missing terms to be of different form and power law dependence. What this means is a matter of future work, rather than speculation. We do not impute a failure of two-loop terms but suggest a cause may lie in two-photon diagrams, or something far more intriguing.

In medium-\emph{Z} ion transitions, a prior estimate of 0.1 eV was made for the magnitude of missing correlation effects to the QED contributions~\cite{Artemyev}. This may result in approximately 20 ppm uncertainty or miscalculation of atomic transition energies \cite{Persson96}. A recent discussion has highlighted the importance of further investigation of the experimental discrepancies~\cite{Lindgren2013}. In the latter development, specific uncalculated terms of order $Z^3$ are predicted.
Theory predicts that the first order QED corrections (vacuum polarization and self-energy) should scale as approximately $Z^4$. Higher order Feynman diagrams generally include $log$ terms or additional powers of $Z\alpha^2$. Correlation terms especially for two-electron systems generally involve shielding or coupling which scales as powers of $(1/Z)$, so the order of the polynomial $n\simeq 4$ is a reasonable hypothesis.

Interestingly (GWF Drake, private communication), the leading term not included correctly by the unified
method is of order $\alpha^4 Z^4$, in harmony with the $Z^4$ scaling that we find for the
discrepancy.  This can be understood as follows.  The second-order relativistic corrections
are of leading orders $\alpha^4 Z^6$ and $\alpha^4 Z^5$ (in units of Ry).  Both of these are fully
accounted for by the unified method.  The leading term not accounted for is therefore of order
$\alpha^4 Z^4$, which is the next term in the $1/Z$ expansion of the higher-order
relativistic corrections.

\section{Conclusions and Outlook to the Future}

Techniques based on the refinement of the understanding of Bragg diffraction over the past several decades, coupled with the development of new methods of producing highly charged ions, currently enable the measurement of the x-ray spectra of highly charged ions to an accuracy of approximately 10 parts-per-million (ppm).  Bragg-based measurements at the 1 ppm level raise numerous systematic errors (corrections due to $n\ge 3$ satellites, depth penetration of wavefields, detector registration and curvature) that have not been adequately addressed in published results to date, and others of less well-defined magnitude for the case of highly charged ions excited by electron collisions (quantum interference shifts)~\cite{Hessels,Sansonetti}.

The quest for high accuracy absolute spectroscopy in the low-\emph{Z} ultraviolet-visible regime has advanced technology and provided powerful new tools for research and applications, yielding Nobel Prizes recently to H\"ansch~\cite{Udem}.
and to Wieman, Cornell and Ketterle~\cite{Bennett}. 
The work of H\"ansch in particular has pushed the uncertainty of the 1S -- 2S hydrogen transition energy to  $4.2$ parts in $10^{15}$~\cite{Hansch1s2s2011}. Core transitions in medium-$Z$ systems cannot be addressed by conventional lasers, so new techniques need to be developed, as discussed here. Efforts in medium-\emph{Z} systems are limited by available calibration techniques (typically to 15 ppm or so), but with lower fractional accuracy can probe similar physics due to the enhancement of contributions~\cite{Beiersdorfer2010} (Fig.~\ref{fig:nature_figure3Ea}). 
A recent paper has provided evidence for a functional discrepancy of measurements of the helium-like $w$ line from advanced QED theory~\cite{Chantler2012PRL}. Here we find that this discrepancy also appears in a series of transitions ($x$, $y$, $z$ in addition to $w$), strengthening the significance of our original finding and allowing the possible origin of the discrepancy in the upper or lower levels to be investigated. Of course experimental limitations can be responsible for any discrepancy, but the pattern and independence of the result with data from different sources and groups helps to minimise this possibility. Work from our (and other) groups reported at recent conferences suggest that the trend of discrepancy with theory may be larger (or smaller) than presented here. We encourage more independent research in theory and experiment, and especially recommend blind protocols for analysis.

Our result can be directly related to $Z^4$ or $Z^6 (1/Z)$ expansion terms, for example, or to combinations with multiple terms. In other words, this is suggestive evidence that specific Feynman diagrams involving virtual photons and electron correlation are not fully accounted for; or of course additional physics with a similar dependence may be present.

\bibliographystyle{unsrt}

\end{article}

\begin{table}
\caption{Experimental weighted mean values used in this work.\label{table_dataWM}}
\begin{tabular}{@{\vrule height 10.5pt depth4pt  width0pt} r | r | r}
$Z$ & discrepancy (eV) & $\sigma $ (eV) \\ \hline
16 &     0.011624 &     0.0285 \\
18 `Fit A' &    -0.01471 &     0.0198 \\
18 `Fit B' &     0.00868 &     0.0072 \\
18 `Fit C' &     0.00209 &     0.0041 \\
19 &     0.16549 &     0.119 \\
20 &     0.04957 &     0.186 \\
21 &     0.12842 &     0.151 \\
22 &     0.13563 &     0.0415 \\
23 &     0.00034 &     0.0736 \\
24 &     0.37393 &     0.316 \\
26 &     0.23567 &     0.1089 \\
27 &    -0.46935 &     0.632 \\
32 &     0.42317 &     0.1171 \\
36 &     0.47794 &     0.1298 \\
54 &     1.1570 &     1.080 \\
92 &    10.8261 &    29.04 \\
\end{tabular}
\end{table}


\begin{table}
\caption{Experimental values used in this work.\label{table_data}}
\begin{tabular} {c c c c c c c}
\emph{Z} & element & line & energy (eV) & $\sigma$ (eV) & discrepancy (eV) & reference \\ \hline
16 &  S & w &   2461.2735 &  0.4886 &  0.6443 &    \cite{Aglitskii1974} \\ 
16 &  S & w &   2460.6874 &  0.1465 &  0.0582 &     \cite{Aglitsky} \\ 
16 &  S & w &   2461.8000 &  3.0000 &  1.1708 &        \cite{Cocke1974} \\ 
16 &  S & w &   2460.6410 &  0.0320 &  0.0118 &      \cite{Kubicek2009} \\ 
16 &  S & w &   2460.6700 &  0.0900 &  0.0408 & \cite{schleinkofer} \\ 
16 &  S & y &   2447.0500 &  0.1100 & -0.0939 & \cite{schleinkofer} \\ 
18 & Ar & w &   3139.6000 &  0.2500 &  0.0179 &       \cite{Briand1983} \\ 
18 & Ar & w &   3139.5830 &  0.0630 &  0.0009 &       \cite{Bruhns2007} \\ 
18 & Ar & w &   3139.5517 &  0.0366 & -0.0304 &    \cite{DeslattesBeyer} \\ 
18 & Ar & w &   3140.1000 &  0.7000 &  0.5179 &      \cite{Dohmann1979} \\ 
18 & Ar & w &   3138.9000 &  0.9000 & -0.6821 &      \cite{Neupert1971} \\ 
18 & Ar & x &   3126.2830 &  0.0363 & -0.0066 &    \cite{DeslattesBeyer} \\ 
18 & Ar & y &   3123.5208 &  0.0362 & -0.0136 &    \cite{DeslattesBeyer} \\ 
19 &  K & w &   3511.4048 &  0.4972 &  0.9432 &    \cite{Aglitskii1974} \\ 
19 &  K & w &   3510.5796 &  0.1229 &  0.1180 & \cite{Beiersdorfer89} \\ 
20 & Ca & w &   3902.4273 &  0.1860 &  0.0496 &            \cite{Aglitsky} \\ 
21 & Sc & w &   4315.5408 &  0.1510 &  0.1284 & \cite{Beiersdorfer89} \\ 
22 & Ti & w &   4750.1702 &  0.9100 &  0.5261 &    \cite{Aglitskii1974} \\ 
22 & Ti & w &   4749.7335 &  0.1662 &  0.0894 & \cite{Beiersdorfer89} \\ 
22 & Ti & w &   4749.8520 &  0.0720 &  0.2079 & \cite{Chantler2012PRL} \\ 
22 & Ti & x &   4733.8335 &  0.1311 &  0.0327 &       \cite{PayneJPB2014} \\ 
22 & Ti & y &   4727.0667 &  0.1000 &  0.1294 &       \cite{PayneJPB2014} \\ 
22 & Ti & z &   4702.0782 &  0.0723 &  0.1036 &        \cite{PayneJPB2014} \\ 
23 &  V & w &   5204.3904 &  1.0923 & -0.7749 &    \cite{Aglitskii1974} \\ 
23 &  V & w &   5205.5922 &  0.5464 &  0.4269 &     \cite{Aglitsky} \\ 
23 &  V & w &   5205.2644 &  0.2082 &  0.0991 & \cite{Beiersdorfer89} \\ 
23 &  V & w &   5205.1000 &  0.1400 & -0.0653 &     \cite{CTC00} \\ 
23 &  V & x &   5189.1200 &  0.2200 &  0.3822 &     \cite{CTC00} \\ 
23 &  V & y &   5180.2200 &  0.1700 & -0.1064 &      \cite{CTC00} \\ 
23 &  V & z &   5153.8200 &  0.1400 & -0.0762 &      \cite{CTC00} \\ 
24 & Cr & w &   5682.6562 &  0.5209 &  0.5878 &     \cite{Aglitsky} \\ 
24 & Cr & w &   5682.3176 &  0.3978 &  0.2492 & \cite{Beiersdorfer89} \\ 
26 & Fe & w &   6700.7617 &  0.3621 &  0.3270 &    \cite{Aglitsky} \\ 
26 & Fe & w &   6700.7254 &  0.2010 &  0.2907 & \cite{Beiersdorfer89} \\ 
26 & Fe & w &   6700.9000 &  0.2680 &  0.4653 &       \cite{Briand1984} \\ 
26 & Fe & w &   6700.4025 &  0.3172 & -0.0322 &      \cite{CTC07} \\ 
26 & Fe & x &   6682.7000 &  0.2673 &  0.3661 &       \cite{Briand1984} \\ 
26 & Fe & y &   6667.5000 &  0.2667 & -0.0786 &       \cite{Briand1984} \\ 
27 & Co & w &   7241.6439 &  0.6323 & -0.4694 &            \cite{Aglitsky} \\ 
32 & Ge & w &  10280.3573 &  0.2715 &  0.1398 &     \cite{Germanium} \\ 
32 & Ge & w &  10280.7000 &  0.2200 &  0.4825 &     \cite{MacLaren1992} \\ 
32 & Ge & x &  10259.5155 &  0.3693 &  0.6416 &     \cite{MacLaren1992} \\ 
32 & Ge & y &  10220.9316 &  0.2275 &  0.1320 &     \cite{Germanium} \\ 
32 & Ge & y &  10221.7911 &  0.3475 &  0.9915 &     \cite{MacLaren1992} \\ 
32 & Ge & z &  10181.3324 &  0.5192 &  0.9456 &     \cite{MacLaren1992} \\ 
36 & Kr & w &  13113.8000 &  1.2000 & -0.6705 &      \cite{Briand84} \\ 
36 & Kr & w &  13115.4500 &  0.3000 &  0.9795 &   \cite{Indelicato86} \\ 
36 & Kr & w &  13114.7800 &  0.7100 &  0.3095 &      \cite{Widmann} \\ 
36 & Kr & w &  13114.6800 &  0.3600 &  0.2095 &      \cite{Widmann1996} \\ 
36 & Kr & x &  13091.5300 &  0.8200 &  0.6643 &      \cite{Widmann} \\ 
36 & Kr & x &  13091.1700 &  0.3700 &  0.3043 &      \cite{Widmann1996} \\ 
36 & Kr & y &  13026.8000 &  0.3000 &  0.6835 &   \cite{Indelicato86} \\ 
36 & Kr & y &  13026.3000 &  0.7100 &  0.1835 &      \cite{Widmann} \\ 
36 & Kr & y &  13026.2900 &  0.3600 &  0.1735 &      \cite{Widmann1996} \\ 
36 & Kr & z &  12979.7200 &  0.6600 &  0.4544 &      \cite{Widmann} \\ 
36 & Kr & z &  12979.6300 &  0.4100 &  0.3644 &      \cite{Widmann1996} \\ 
54 & Xe & w &  30629.1000 &  3.5000 & -0.9512 &      \cite{Briand1989} \\ 
54 & Xe & w &  30631.2000 &  1.2000 &  1.1488 &        \cite{Thorn2009} \\ 
54 & Xe & y &  30209.6000 &  3.5000 &  3.3348 &       \cite{Briand1989} \\ 
92 &  U & w & 100626.0000 & 35.0000 & 15.1100 &       \cite{Briand1990} \\ 
92 &  U & y &  96171.0000 & 52.0000 &  1.3700 &       \cite{Briand1990} \\ 
\hline
\end{tabular}
\end{table}

\end{document}